\documentclass[12pt]{article}

\usepackage{epsfig,graphicx,bm,amsmath,amsfonts}
\usepackage{epstopdf}

\newcommand{\mc}[1]{\mathcal{#1}}

\begin{document}

\setlength{\unitlength}{1mm}

\begin{titlepage}

\begin{flushright}
AEI-2009-092
\end{flushright}
\vspace{2cm}

\begin{center}
{\bf \Large Quasilocal formalism and thermodynamics of asymptotically flat black objects}\\
\vspace*{2cm}
\end{center}

\vspace{1cm}

\begin{center}
\bf{Dumitru Astefanesei$^{a}$,}
\bf{Robert B. Mann$^{b,c}$}
\bf{Maria J. Rodriguez$^{a}$}
\bf{Cristian Stelea$^{d}$}

\vspace{.5cm}

{\small \it $^{a}$ Max-Planck-Institut f\"ur Gravitationsphysik, Albert-Einstein-Institut, 14476 Golm, Germany}\\
\vspace{2mm}
{\small \it $^{b}$Perimeter Institute for Theoretical Physics, Ontario N2J 2W9, Canada}\\
\vspace{2mm}
{\small \it $^{c}$ Department of Physics, University of Waterloo Waterloo, Ontario N2L 3G1, Canada}\\
\vspace{2mm}
{\small \it $^{d}$ Department of Physics and Astronomy, University of British Columbia, Vancouver, Canada}\\[.5em]

{\small {\tt dumitru@aei.mpg.de, mann@avatar.uwaterloo.ca,\\ maria.rodriguez@aei.mpg.de, stelea@phas.ubc.ca}}

\end{center}

\vspace{1cm} 

\begin{abstract}
We study the properties of 5-dimensional
black objects by using the {\it renormalized} boundary
stress-tensor for locally asymptotically flat spacetimes. 
This provides a more refined form of the quasilocal formalism 
which is useful for a holographic interpretation of 
asymptotically flat gravity. We apply this technique to 
examine the thermodynamic properties of black holes, black 
rings, and black strings. The advantage of using this method 
is that we can go beyond the `thin ring' approximation and 
compute the boundary stress tensor for any general (thin or 
fat) black ring solution. We argue that the boundary stress 
tensor encodes the necessarily information to distinguish 
between black objects with different horizon topologies in 
the bulk. We also study in detail the susy black ring 
and clarify the relation between the asymptotic charges and 
the charges defined at the horizon. Furthermore, we obtain 
the balance condition for `thin' dipole black rings.
\end{abstract}

\vspace{2cm} 


\end{titlepage}

\tableofcontents

\section{Introduction}
A remarkable development in theoretical physics was the
discovery of a close relationship between the laws of thermodynamics
and certain laws of black hole physics. The black hole represents
the equilibrium end state of gravitational collapse, so on general
grounds we might expect it to be the state of maximum entropy for
a self-gravitating system. The relationship between thermodynamic
entropy and the area of an event horizon is one of the most robust
and surprising results in gravitational physics.

In a very basic sense, gravitational entropy can be regarded as
arising from the Gibbs-Duhem relation applied to the path-integral
formulation of quantum gravity \cite{Mann:2002fg}. In the
semiclassical limit this yields a relationship between gravitational
entropy and other relevant thermodynamic quantities, such as mass,
angular momentum, and other conserved charges.

This relationship was
first explored in the context of black holes by Gibbons and Hawking
\cite{Gibbons:1976ue}, who argued that the thermodynamical potential
is equal to the Euclidean gravitational action multiplied by the temperature. In
this approach, the partition function for the gravitational field is
defined by a sum over all smooth Euclidean geometries with a period
$\beta$ in imaginary time. The integral is computed by using the saddle
point approximation.

When applying this
method to Schwarzschild black hole, the calculation is purely
gravitational (no additional `matter' fields are present) and the entropy
is one-fourth of the horizon area. Therefore, this result confirms that the entropy
is an {\it intrinsic} property of black holes.

It is well known that, due to the equivalence principle, a local
definition of energy in gravity theories is meaningless. One of
the most fruitful approaches in computing conserved quantities
has been to employ the quasilocal formalism \cite{Brown:1992br}.
The basic idea of Brown and York was to define a {\it quasilocal}
energy. That is to enclose a given region of spacetime with some surface, and
to compute the energy\footnote{In fact, one can compute all
relevant thermodynamic quantities \cite{Jolien,CreightonMann,BoothMann}.} with respect
to that surface.

For an asymptotically flat spacetime, it is possible to extend the
quasilocal surface to {\it spatial} infinity without difficulty, provided
one incorporates appropriate boundary terms (counterterms) in the
action to remove divergences \cite{Lau:1999dp, Mann:1999pc, Kraus:1999di}. This
method was inspired by the holographic renormalization method in AdS spacetimes
\cite{skenderis} (see, e.g., \cite{Astefanesei:2008wz, Cai:1999xg} for counterterms
in more general theories) and the counterterms were obtained by considering the
flat space limit (the AdS radius is infinite).

Subsequently, the authors of  ref. \cite{Astefanesei:2005ad} proposed a
{\it renormalized} stress-tensor for a general class of
stationary spacetimes which are locally asymptotic to flat
space --- it was computed by varying the total
action (including the counterterms) with respect to the boundary
metric. The conserved quantities can be constructed from this stress
tensor via the algorithm of Brown and York \cite{Brown:1992br}. As an example,
this method was applied in ref. \cite{Astefanesei:2005ad} to understand the
thermodynamics of the dipole ring \cite{Emparan:2004wy}.

However, there are subtleties in taking the flat spacetime limit
and the references
\cite{Lau:1999dp, Mann:1999pc, Kraus:1999di} did not present a rigorous
justification for considering these counterterms.\footnote{One
problem with the flat spacetime is that its holographic description
seems to be nonlocal \cite{Marolf:2006bk}.}

In flat spacetime the usual gravity covariant action supplemented
with the boundary Gibbons-Hawking term does not satisfy a valid
variational principle. Mann and Marolf have constructed a valid
covariant variational principle by adding an appropriate local
boundary term \cite{Mann:2005yr} (see, also, \cite{Mann:2006bd,
Astefanesei:2006zd}). This
counterterm makes direct contact with the background subtraction
procedure. They have also demonstrated that the conserved quantities
related to the boundary stress-tensor agree with the usual $ADM$
definitions\footnote{The conserved quantities defined in this
way also generalize the usual definitions to allow, e.g.,
non-vanishing NUT charge in four-dimensions --- see, also,
\cite{Liko:2008rr} for a different approach to compute the
NUT charge.}
\cite{Arnowitt:1962hi} (see, also, \cite{Baskaran:2003pk}).
In particular, this work provides a rigorous justification for
the proposal of the renormalized stress tensor of \cite{Astefanesei:2005ad}.

In the asymptotically flat case, the only neutral static black hole
is the five dimensional Schwarzschild-Tangherlini solution \cite{Gibbons:2002av}.
The rotating case is more involved and includes both Myers-Perry
black holes \cite{Myers:1986un} and black rings
\cite{Emparan:2001wn, Emparan:2006mm}.

In this paper we apply the method of \cite{Astefanesei:2005ad}
in a systematic way to study the thermodynamics of asymptotically
flat black objects. We will restrict our considerations to
five dimensions, although a similar formalism should be valid for any
spacetime dimension (see \cite{Astefanesei:2006zd} for a similar
analysis in four dimensions).

 At this point we pause to explain the advantages of using 
this method. First, computations of the ADM stress tensor 
have been carried out just in the `thin ring' approximation (see 
\cite{Emparan:2006mm} and the references therein).   In this paper
we go beyond this approximation, computing 
the complete boundary stress tensor for any (fat or thin) black 
ring solution. 

Furthermore, while there is a computation of the 
asymptotic charges within the ADM formalism, there is no computation 
of the action. Indeed, to the best of our knowledge, there is no known background 
subtraction calculation for black rings. In this sense, analysis 
of thermodynamics in the grand canonical ensemble that is presently found in the
literature is incomplete. However, we explicitly compute the thermodynamic 
potential and recover some of the previous results. For concreteness, 
we also provide the associated phase diagrams based on the Gibbs 
potential --- these plots have not before appeared in the literature. 

Another noteworthy application of our work is in understanding how the quasilocal 
formalism applies to theories with a Chern-Simons term. We present a detailed 
study of the susy black ring, which is helpful for clarifying the relation 
between the asymptotic charges and the charges computed at the horizon in 
this case.

An outstanding question concerning black rings
is how an asymptotic observer can distinguish between a black ring and 
a black hole with the same asymptotic charges. In this paper, our modest 
purpose is merely to point out some tools to address this situation: we argue 
that the required information is encoded in the quasilocal stress tensor.

Due to the significant promise the quasilocal formalism (supplemented 
with counterterms) has for further applications, we have designed our paper to be
self-contained,  with concrete examples --- it can be also considered 
as an introduction to the subject.

The remainder of this paper is organized as follows. Section 2
contains a review of the results of \cite{Brown:1992br, Astefanesei:2005ad}
and also the complex instanton method \cite{Brown:1990di,Brown:1990fk}.
 In Section 3, we investigate in detail the thermodynamic  properties
of neutral black ring (with one angular momentum) and black hole in the 
grand canonical ensemble. In Section 4 we examine a few charged black objects
including the susy black ring. In particular, we present a discussion
of the balance condition for the {\it thin} dipole-charged black rings 
within the quasilocal formalism. Section
5 closes with a comprehensive discussion and some observations about
our results.

\section{General method}
In this section we review the basic framework that we will use to
study the thermodynamics of asymptotically flat black objects. First,
we present an overview of the quasilocal formalism and counterterm
method. Then, we discuss the complex instanton method and the role
of the quasi-Euclidean section in understanding  black ring thermodynamics.

\subsection{Quasilocal formalism}
The action functional for general relativity contains a contribution
$I_G[g]$ from the gravitational field $g_{\mu\nu}$ and a
contribution $I_M[\Psi;g]$ from the matter fields, which we
collectively denote $\Psi$. In the early days of studying the path
integral for gravity, the gravitational action for some region $M$
was written as a sum of a Hilbert term $I_H[g]$, a term evaluated on
its boundary $\partial M$, $I_B[g]$, and a {\it nondynamical} term
$I_{ref}[g_{ref}]$:
\begin{eqnarray}
I_G[g] &\equiv& I_H[g]+I_B[g]-I_{ref}[g_{ref}] \nonumber\\
&=& \frac{1}{16\pi
G}\int_MR\,\sqrt{-g}\,d^{5}x+\frac{\epsilon}{8\pi G}\int_{\partial
M} (K-K_0)\sqrt{-h}\,d^{4}x\, .
\label{generalI}
\end{eqnarray}

Here, $K$ is the extrinsic curvature of $\partial M$, $\epsilon$ is
equal to $+1$ where $\partial M$ is timelike and $-1$ where
$\partial M$ is spacelike, and $h$ is the determinant of the induced
metric on $\partial M$.

The existence of a boundary term in the
gravitational action is an atypical feature of field theories ---
 it appears due to the fact that $R$, the gravitational Lagrangian
density, contains second derivatives of the metric tensor. This term
is required so that upon employing the variational principle with  metric
variations fixed at the boundary, the action yields the Einstein equations.

Let us elucidate now the role of $I_{ref}$. Clearly it affects the
numerical value of the action but not the equations of motion. The
main observation is that even at tree-level, the gravitational
action contains divergences that arise from integrating over the
infinite volume of spacetime. Hence one should regularize the action
to get finite results.

One way to do this is by subtracting a new term
$I_{ref}[g_{ref},\Psi_{ref}]$ from the action \cite{Brown:1992br}. The action and
conserved quantities are calculated with respect to this {\it
reference} spacetime which is interpreted as the ground state of the
system. An important difficulty with this approach is that it is not
always possible to embed a boundary with a given induced metric into
the reference background \cite{CCM}.

Fortunately, there is a second way \cite{Mann:1999pc,
Kraus:1999di,Astefanesei:2005ad, Mann:2005yr}
to regularize the gravitational action and the
stress-energy of gravity. Namely, one supplements the quasilocal
formalism of Brown and York \cite{Brown:1992br} by including boundary
counterterms. This method was inspired from the stringy AdS/CFT
correspondence \cite{skenderis}, where the infrared
divergences of the gravity in the bulk (due to integration over
infinite volumes) are dual to ultraviolet divergences in
the dual boundary conformal field theory. These divergences can be
removed by adding additional boundary terms that are geometric
invariants of the induced boundary metric, leading to a finite
total action.

The counterterms are built up by curvature invariants of the
boundary $\partial M$ (which is sent to infinity after the
integration) and thus, obviously, they do not alter the bulk
equations of motion. Rather than employ the counterterm proposal in
ref \cite{Mann:2005yr}, for asymptotically flat solutions (on the
Euclidean section)\footnote{The action is computed on the Euclidean
section but the stress tensor can be computed on the Lorentzian section.}
we consider the following counterterm expression
\begin{eqnarray}
I_{ct}[h]=-\frac{c}{8 \pi G}\int_{\partial M} d^{4} x \sqrt{-h}
\sqrt{\mathcal{R}}\,\,\, ,
\label{Ict}
\end{eqnarray}
where $\mathcal{R}$ is the Ricci scalar of the induced metric on the
boundary $h_{ij}$. The constant $c$ that enters the above relation
depends on the boundary topology --- one finds $c=\sqrt{3/2}$ for a
boundary topology $S^3\times S^1$ and $c=\sqrt{2}$ for a
$S^2\times R\times S^1$ topology.  This choice of  boundary term
yields an action that is stationary on solutions, so long as the
spatial cut-off induces a boundary of the form
$S^n \times {\mathbb R}^{d-n-1}$ \cite{Mann:2005yr}.

Varying the total action, $I=I_{H}[g]+I_B[g]+I_{ct}[h]$, with respect to the
boundary metric $h_{ij}$, we compute the divergence-free boundary
stress tensor \cite{Astefanesei:2005ad}
\begin{eqnarray}
\tau_{ij}\equiv\frac{2}{\sqrt{-h}}\frac{\delta I}{\delta h^{ij}}=
\frac{1}{8\pi G}\Big( K_{ij}-h_{ij}K
-\Psi(\mathcal{R}_{ij}-\mathcal{R}\,h_{ij})-h_{ij}\square\Psi+\Psi_{;ij}\Big), \label{Tik}
\end{eqnarray}
where $\Psi=\frac{c}{\sqrt{\mathcal{R}}}$.\\

The boundary metric can be written, at least locally, in ADM-like
form
\begin{eqnarray}
 h_{ij}dx^i dx^j=-N^2\,dt^2+\sigma_{a b}\,(dy^a
+N^a\,dt)(dy^b +N^b\,dt),
 \label{sigma}
\end{eqnarray}
where $N$ and $N^a$ are the lapse function and the shift vector
respectively and the ${y^a}$ are the intrinsic coordinates on the
closed surfaces $\Sigma$.

Provided the boundary geometry has an isometry generated by a
Killing vector $\xi ^{i}$, a conserved charge
\begin{eqnarray}
 {\mathfrak Q}_{\xi }=\oint_{\Sigma }d^3y\,
\sqrt{\sigma}n^i\,\tau_{ij}\,\xi^j, \label{charge}
\end{eqnarray}
can be associated with a closed surface $\Sigma $ (with normal
$n^{i}$). Physically this means that a collection of observers on
the hypersurface whose metric is $h_{ij}$ all observe the same value
of ${\mathfrak Q}_{\xi }$ provided this surface has an isometry
generated by $\xi^{i}$ \cite{BoothMann,Booth:1998pb}. For example, if $%
\xi =\partial /\partial t$ then ${\mathfrak Q}$ is the conserved mass/energy $%
{\mathfrak M}$.

One of the appealing features of this approach is that
it provides elegant `natural' definitions of quasilocal
energy and angular momentum.

Gravitational thermodynamics is then formulated via the Euclidean
path integral
\[
Z=\int D\left[ g\right] D\left[ \Psi \right] e^{-I\left[ g,\Psi
\right] }\simeq e^{-I},
\]%
where one integrates over all metrics and matter fields between some
given initial and final Euclidean hypersurfaces, taking $\tau $ to
have some period $\beta $. The period is determined by requiring the
Euclidean section be free of conical singularities. Semiclassically,
the total action is evaluated from the classical solution of field
equations, yielding an expression for the entropy
\begin{eqnarray}
S=\beta ({\mathfrak M}-\mu _{i}{\mathfrak C}_{i})-I,  \label{GibbsDuhem}
\end{eqnarray}%
upon application of the Gibbs-Duhem relation to the partition function \cite%
{Mann:2002fg} (with chemical potentials ${\mathfrak C}_{i}$ and
conserved charges $\mu _{i}$). The first law of thermodynamics is then
\begin{eqnarray}
dS=\beta (d{\mathfrak M}-\mu _{i}d{\mathfrak C}_{i}).  \label{1stlaw}
\end{eqnarray}
%

\subsection{Complex instanton}
The thermodynamic properties of a dipole black ring were derived
by using the counterterm method \cite{Astefanesei:2005ad}. Also,
using the Gibbs-Duhem relation, a non-trivial check
of the entropy/area relationship for the dipole ring was obtained.

However, a key point regarding one's intuition about the Euclidean section
does not apply to black rings. Naively, one expects to find
a real Euclidean section for a black ring solution. However it was shown
in \cite{Astefanesei:2005ad, Elvang:2006dd} that the situation is
more subtle: there is no real {\it non-singular} Euclidean section in this case.
Nevertheless, as argued in \cite{Brown:1990fk}, these configurations
still can be described by a complex geometry  and a real
action\footnote{This method was also used in \cite{Booth:1998pb}.}.

As in \cite{Astefanesei:2005ad}, we adopt the `quasi-Euclidean' method of
\cite{Brown:1990fk} in which the Wick transformations
affect the intensive variables, such as the lapse and shift
($N\rightarrow -iN$ and $N^k\rightarrow -iN^k$), but for which
the extensive variables (such as energy) remain invariant. It is
worth mentioning that the Cauchy data and the
equations of motion remain invariant under this
complexification.

Now, let us recapitulate the general formalism from
\cite{ Brown:1990di,Brown:1990fk}. We begin with the
standard ADM-decomposition: first, select an {\it arbitrary}
foliation of spacetime by specifying the lapse function $N$ and
the shift vector $N^a$. Defining $\gamma_{ij}$ to be the induced
metric on the spacelike hypersurfaces of constant time, the {\it
full} spacetime metric is given by:
\begin{eqnarray} ds^2=g_{\mu\nu}\,dx^\mu dx^\nu=-N^2\,dt^2 +
\gamma_{ij}\,(dx^i+N^i\,dt)(dy^j+N^j\,dt) \, . \label{ADM} \end{eqnarray}
Next we choose initial values for the tensor fields
$\gamma_{ij}$ and $K_{ij}$, where $K_{ij}$ is the extrinsic
curvature of the spacelike hypersurfaces. The initial values must be
solutions of the $constraint$ equations and so the choice is not
entirely arbitrary. Then, the appropriate complexification that
preserves the constraints and the dynamical equations of motion of
{\it stationary} spacetimes is given by replacing $N$ with $-iN$ and
also changing the shift vector $N^i$ and the gauge potential $A_0$ from
real to imaginary. The complex Euclidean metric becomes
\begin{eqnarray} ds^2=N^2\,d\tau^2 +
\gamma_{ij}\,(dy^i-i\,N^i\,d\tau)(dy^j-i\,N^j\,d\tau) \, .
\label{pseudoeuclidean} \end{eqnarray}
The key point is that the energy, angular momentum, and electric
charge are defined by surface integrals of the Cauchy data and so
they remain {\it real} with their physical values.

Armed with this formalism, we will be able to investigate the
thermodynamics of black rings. To check consistency, we shall also apply
this method to other examples.

\subsection{Temperature and angular velocity}
An asymptotically flat spacetime is {\it stationary} if and
only if there exists a Killing vector field, $\xi$, that is
timelike near spatial infinity --- it can be normalized such
that $\xi^2\rightarrow -1$. It has been shown that
stationarity implies {\it axisymmetry} \cite{Hollands:2006rj}
and so the event horizon is a Killing horizon.

The general stationary metric\footnote{We use the conventions and
compute the temperature and the angular velocity as in \cite{Astefanesei:2007bf}.}
with an `axial' vector Killing,
$\frac{\partial}{\partial \phi}$ , can be written as
\begin{equation}
ds^2=g_{tt}(\vec{x})dt^2 + 2g_{t\phi}(\vec{x})dt\,dx^\phi+g_{ij}(\vec{x})dx^i\,dx^j.
\label{metric2}
\end{equation}
A stationary spacetime is static, at least near spatial
infinity, if it is also invariant under time-reversal
(i.e., $g_{t\phi}(\vec{x})=0$).

We rewrite the metric (\ref{metric2}) in the ADM form (\ref{ADM}), and so we
obtain:
\begin{equation}\label{lapse}
N^2=\frac{(g_{t\phi})^2}{g_{\phi\phi}}-g_{tt}, \,\,\,\,
N^{\phi}=\frac{g_{t\phi}}{g_{\phi\phi}}, \,\,\,\, \gamma_{ij}=g_{ij}.
\end{equation}
The shift vector evaluated at the horizon reproduces the angular velocity
of the horizon:
\begin{equation}
\label{angularvelocity}
\Omega^{\phi}_H=-\left.N^{\phi}\right\vert_H =
-\left.\frac{g_{t\phi}}{g_{\phi\phi}}\right\vert_H\,.
\end{equation}
To compute the temperature, we should eliminate the conical singularity
in the $(\tau,r)$ sector. Let us define a new radial coordinate
$R=\sqrt{N^2}$. Thus we have $dR=\frac{1}{2}(N^2)^{-1/2}(N^2)'dr$, and
we get
$$
ds^2=N^2d\tau^2+g_{rr}dr^2=g_{rr}\frac{4N^2}{[(N^2)']^2}\left[
dR^2+\frac{[(N^2)']^2R^2}{4N^2g_{rr}}d\tau^2\right].
$$
Hence, in the vicinity of $r=r_H$, we see that $R=0$ is like the origin of
the polar coordinates provided that we identify $\tau$ with period $\Delta\tau$
given by
$$
\left.\frac{(N^2)'}{2\sqrt{N^2g_{rr}}}\right\vert_H\Delta\tau=2\pi,
$$
and so the temperature is
\begin{eqnarray}
\label{temp}
T=\frac{1}{\Delta\tau}=\left.\frac{(N^2)'}{4\pi\sqrt{N^2g_{rr}}}\right\vert_H\,.
\end{eqnarray}

The lapse $N$ and shift $N^{\phi}$are not dynamical quantities -- they can be
specified freely and correspond to the arbitrary choice of coordinates. It is
important to emphasize that the lapse determines the slicing of spacetime and
the choice of shift vector determines the spatial coordinates.

Note that, with this foliation of spacetime, the black hole horizon is at $N^2=0$.

\section{Vacuum solutions}
In this section we apply the counterterm method to five-dimensional
vacuum solutions of Einstein gravity. We explicitly show how to compute
the action and the conserved charges for the Myers-Perry black hole and for
the black ring. By using the action computed on the quasi-Euclidean section
we also present a detailed analysis of the thermodynamic stability in
canonical and grand canonical ensembles.

\subsection{The model}
The existence of non-spherical horizon topologies in dimensions higher
than four implies that the notion of
black hole uniqueness is very much weaker in higher dimensions. In fact,
the existence of a black ring with the same conserved charges as the
black hole is a counterexample to a straightforward extension of the 4-dimensional
black hole uniqueness theorems.

We start by discussing the spinning vacuum solutions of Einstein field
equations: the black hole \cite{Myers:1986un} and the black ring
\cite{Emparan:2001wn} --- a detailed discussion of black ring physics
can be found in \cite{Emparan:2006mm}.

Using the conventions in \cite{Elvang:2003mj} we can write a general
line element representing both solutions as follows
\begin{eqnarray}
\label{generalmetricunbalanced}
  ds^2 &=& -\frac{F(x)}{F(y)} \left(dt+
     R\sqrt{\lambda\,\nu} \,(1 + y)\, d\psi\right)^2  \label{ring0}\\
  &&
   +\frac{R^2}{(x-y)^2}
   \left[ -F(x) \left( G(y)\, d\psi^2 +
   \frac{F(y)}{G(y)}\, dy^2 \right)
   + F(y)^2 \left( \frac{dx^2}{G(x)}
   + \frac{G(x)}{F(x)}\,d\phi^2\right)\right]  \nonumber
\end{eqnarray}
with
\begin{eqnarray}
  F(\xi) = 1 - \lambda\xi \, ,
\qquad  G(\xi) = (1 - \xi^2)(1-\nu \xi) \, .
\end{eqnarray}
$R,\lambda$ and $\nu$ are parameters whose appropriate combinations
give the mass and angular momentum.

The variables $x$ and $y$ take values in
\begin{eqnarray}
-1\leq x\leq 1\,,\qquad  -\infty<y\leq-1\,,\quad \lambda^{-1}<y<\infty\,.
\label{xyrange}\end{eqnarray}
As shown in \cite{Elvang:2003mj}, in order to balance forces in the
ring one must identify
$\psi$ and $\phi$ with equal period
\begin{eqnarray}
\Delta\phi=\Delta\psi=\frac{4\pi \sqrt{F(-1)}}{|G'(-1)|}=
\frac{2\pi\sqrt{1+\lambda}}{1+\nu}\,.
\label{phsiperiod}
\end{eqnarray}
This eliminates the conical singularities at the fixed-point sets $y=-1$
and $x=-1$ of the Killing vectors $\partial_\psi$ and $\partial_\phi$,
respectively.

However there still is the possibility of conical singularities at $x=+1$.
These can be avoided in either of two ways. Fixing
\begin{eqnarray}
\lambda=\lambda_c\equiv \frac{2\nu}{1+\nu^2}\qquad {\rm (black\; ring)}
\label{lambring}
\end{eqnarray}
makes the circular orbits of $\partial_\phi$ close off smoothly also at
$x=+1$. Then $(x,\phi)$ parametrize a two-sphere, $\psi$ parametrizes a
circle, and the solution
describes a black ring. Alternatively, if we set
\begin{eqnarray}
\lambda=1\qquad {\rm (black\; hole)}
\label{lambhole}\end{eqnarray}
then the orbits of $\partial_\phi$ do not close at $x=+1$. Then
$(x,\phi,\psi)$ parametrize an $S^3$ at constant $t,y$. The solution is
the same as the spherical black hole of \cite{Myers:1986un} with a single rotation
parameter. Both for black holes and black rings, $|y|=\infty$ is an
ergosurface, $y=1/\nu$ is the event horizon, and the inner, spacelike
singularity is reached as $y\to\lambda^{-1}$ from above.

The parameters $\nu$ and $\lambda$ have the range
\begin{eqnarray}
\label{nurange}
0\leq\nu<\lambda<1\,.
\end{eqnarray}
As $\nu\to 0$ we recover a non-rotating black hole, or a very thin black
ring. At the opposite limit, $\nu\to 1$, both the black hole and the
black ring get flattened along the plane of rotation, and at $\nu=1$
approach the same solution with a naked ring singularity.

 Asymptotic spatial infinity is reached as $x\to y\to-1$.

\subsection{Boundary stress-tensor and conserved charges}
To evaluate asymptotic expressions at spacelike infinity, it is
convenient to introduce coordinates in which the asymptotic flatness
of the solutions becomes manifest.
 Our choice for this transformation is
\begin{eqnarray}
\label{transf} x=1-\frac{2\,\alpha^2\,r^2}{\alpha^2\,r^2+R^2\cos^2 \theta}\,,\qquad
y=1-\frac{2(\alpha^2\,r^2+R^2)}{\alpha^2\,r^2+R^2 \cos^2 \theta}\,,\qquad \alpha=\frac{\sqrt{1+\nu}}{(1+\lambda)}\,,
\end{eqnarray}
$r$  corresponding to a normal coordinate on the boundary, $0\leq
r<\infty$, $0\leq \theta\leq \pi/2$.
In these coordinates, the black ring approaches asymptotically the
Minkowski background
\begin{eqnarray}
ds^2=-dt^2+dr^2+r^2(d\theta^2+\sin^2 \theta\,d\psi^2+\cos^2\theta\,d\phi^2)\,,
\end{eqnarray}
where $\phi$ and $\psi$ are angular coordinates
rescaled according to (\ref{phsiperiod}).

The mass and angular momentum can be computed by employing the quasilocal
formalism and we obtain from eq. (\ref{Tik}) the following relevant boundary
stress-tensor components:
\begin{eqnarray}
\label{emparan-tik}
\tau_{tt}&=&\frac{1}{8\pi G}\left(-\frac{3 R^2  \lambda (1+\lambda) }{1+\nu}\frac{1}{r^3}+F_0\,\frac{\cos 2 \theta}{r^3}+\mc{O}(1/r^5)\right)\,,\nonumber\\
\tau_{t\psi}&\equiv& \tau_{\psi t}=\frac{1}{8\pi G}\left(-\frac{4\,R^3  \sqrt{\lambda  \nu } (1+\lambda )^{5/2}}{(1+\nu )^2}\frac{\sin^2 \theta}{r^3}+\mc{O}(1/r^5)\right)\, ,
\end{eqnarray}
where
\begin{equation}
F_0=-\frac{ R^2 (1+\lambda ) (5+13 \nu - 17\lambda -9 \nu \lambda )}{3 (1+\nu )^2}\, .
\end{equation}
Note that the term in $\tau_{tt}$ containing $F_0$ will make no contribution once integrated over the closed surface $\Sigma$. New non-trivial contributions will appear at subleading order $\mc{O}(1/r^5)$ and correspond to the dipole.

Thus, the  mass and angular momentum of this solution are
\begin{eqnarray}
M=\frac{3\pi R^2}{4G}\frac{\lambda(\lambda+1)}{1+\nu}\,,\qquad
J=\frac{\pi
R^3}{2G}\frac{\sqrt{\lambda\nu}(\lambda+1)^{5/2}}{(1+\nu)^2}\,,
\label{eq:mandj}
\end{eqnarray}
As expected, the mass and angular momentum computed from
the boundary stress tensor according to (\ref{charge}) agree with
the standard ADM expressions \cite{Emparan:2001wn}.

\subsection{Pseudo-Euclidean section and thermodynamic stability}
\label{sec:Stability}
As discussed in Section 2.2, the analytic continuation $t \to -
i\tau$ leads to a complex Euclidean metric.  After performing this action on  the line-element
(\ref{ring0}), the lapse  function is
\begin{eqnarray}
N^2=\frac{F(x)}{F(y)}\,\left[\frac{G(y)F(y)}{\lambda\nu(1+y)^2(x-y)^2-G(y)F(y)}\right]\, ,
\label{lapseBR}
 \end{eqnarray}
and the shift vector is
\begin{eqnarray}
 N^{\psi}=\frac{1}{R}\frac{\sqrt{\lambda\nu}(1+y)(x-y)^2}{\lambda\nu(1+y)^2(x-y)^2-G(y)F(y)}\, .
\label{shift}
\end{eqnarray}
The angular velocity at the horizon reads
\begin{eqnarray}\Omega=\frac{1+\nu}{\sqrt{1+\lambda}}\, \Omega^{\psi}_H=\frac{1}{R}\sqrt{\frac{\nu}{\lambda(1+\lambda)}}\, ,
\label{angularvel}
\end{eqnarray}
where $\Omega^{\psi}_H$ is obtained from eq. (\ref{angularvelocity}).
Using the results in Section $2.3$ it is straightforward to prove that
the temperature and the horizon area are given by
\begin{eqnarray}
\mathcal{A} =
  8 \pi^2 R^3 \frac{\lambda^{1/2} (1+\lambda)(\lambda - \nu)^{3/2}}
       {(1+\nu)^2 (1-\nu)}\,,\quad
  T = \frac{1}{4\pi R}\frac{1-\nu}
  {\lambda^{1/2} (\lambda - \nu)^{1/2} }
     \, .
\label{areatemp}
\end{eqnarray}

Now, we would like to compute the {\it renormalized} action that is
related to the free energy of the system. The scalar curvature $R$
vanishes so only the surface terms give a
contribution to the action. To evaluate these terms, it is
convenient to use the $(r,\theta)$ coordinate system. One finds that
\begin{eqnarray}
\label{KK}
\lim_{r \to \infty}\sqrt{-h}\left(\sqrt{\frac{3}{2}\mathcal{R}}-K\right)&=&\frac{R^2 (1+\lambda ) (\lambda  (1+\nu )-4 (\lambda +\nu +2 \lambda  \nu ) \cos 2 \theta ) \sin 2 \theta }{2 (1+\nu )^2}+\mc{O}(1/r^2)\,,\nonumber
\end{eqnarray}
which is finite. The expression for the total action is
\begin{eqnarray}
\label{emparan-action}
I= \frac{\pi ^2 R^3}{G}\frac{ \lambda ^{3/2} (1+\lambda ) (\lambda -\nu )^{1/2}}{ (1-\nu^2 )}\, .
\end{eqnarray}
It can be verified that
\begin{eqnarray}\label{quasi-Euc-action}
\label{ig} I=\beta (M-\Omega J)-\frac{{\cal A}}{4\,G},
\end{eqnarray}
with $M,~\Omega, J$ and ${\cal A}_H$ given above, while $\beta=1/T$.
Therefore the entropy of this solution is the event
horizon area divided by $4\,G$, as expected. Also,  the first law of thermodynamics,
$dM= T\,dS +\Omega\, dJ$, is satisfied.

To study the phase structure and stability of black objects we must analyze the
potentials and the response functions in different thermodynamic ensembles.
We will briefly describe the potentials, response functions, and stability conditions
in the canonical and grand-canonical ensembles. Previous related
studies can be found in \cite{Astefanesei:2005ad, Elvang:2006dd, Monteiro:2009tc}.

In the grand-canonical ensemble ($i.e.$ for
fixed temperature, angular velocity, and gauge potential), by
using the definition of the Gibbs potential ${\cal G}[T,\Omega]=I/\beta$
and the expression for the angular velocity we obtain
\begin{eqnarray}
\label{Gib} {\cal G}[T,\Omega ]=M-\Omega \,J-T S ,
\end{eqnarray}
As expected, ${\cal G}[T,\Omega]$ is indeed the Legendre
transform of the energy $M[S,J]$ with respect to $S$, $J$ --- the
Legendre transform simply exchanges the role of the variables
associated with the control and response of the system.
A detailed discussion of  thermodynamic stability in different ensembles
is given in \cite{Callen} (we respect the conventions in this book) --- a
nice review of different methods in the context of black hole objects is given
in \cite{Monteiro:2009tc}.

The physical implication of the stability conditions is that they constrain
the response functions of the system. In analogy with the definitions
for thermal expansion in the liquid-gas systems, the specific heat at constant
angular velocity, the isothermal compressibility, and the coefficient of thermal
expansion at the horizon are defined respectively as follows
\begin{eqnarray}
C_\Omega=T\left(\frac{\partial S}{\partial T}\right)_{\Omega}=
-T\left(\frac{\partial^2 {\cal G}}{\partial T^2}\right)_{\Omega}\,,\qquad\epsilon_{T}=\left(\frac{\partial J}{\partial \Omega}\right)_{T}\,,\qquad\alpha=\left(\frac{\partial J}{\partial T}\right)_{\Omega}\,.
\end{eqnarray}
The conditions for the stability of a thermodynamic configuration in the grand
canonical ensemble are
\begin{eqnarray}
C_\Omega>0\,,\qquad\epsilon_{T}>0\,,
\end{eqnarray}
as well as
\begin{eqnarray}\label{detconstrain}
C_{\Omega}\,\epsilon_T-\alpha^2\, T\,>\,0\,.
\end{eqnarray}

On the other hand, when considering a canonical ensemble, the variables are the temperature
T and angular momenta J. The potential is the Helmholtz free energy defined as
\begin{eqnarray}
F[T,~J]=M-TS\,,
\end{eqnarray}
and the entropy is $S=-\left(  \partial F/\partial
T\right) _{J}$. In this case, one finds the following expressions for the response functions for the specific heat at constant angular momentum
\begin{eqnarray}
\label{relcj}
C_J=T\left(\frac{\partial S}{\partial T}\right)_{J} =-T\left(\frac{\partial^2 F}{\partial T^2}\right)_{J}\,,
\end{eqnarray}
and also the inverse of the isothermal compressibility and the coefficient of thermal expansion defined for the grand canonical ensemble. The stability conditions in the canonical ensemble have the same consequences on the constraints on the response functions as in the grand-canonical ensemble. This is due to the equivalence between the heat capacities, that follow from the mathematical relations derived from the basic thermodynamic laws,
\begin{eqnarray}
C_J=C_{\Omega}-T\,\epsilon^{-1}_T\,\alpha^2\,,
\end{eqnarray}
Using (\ref{detconstrain}) one can easily obtain that $C_J\,>0$.

\subsubsection{Black hole}
\begin{figure}
 \centering
  \includegraphics[width=6.9cm,height=4.5cm]{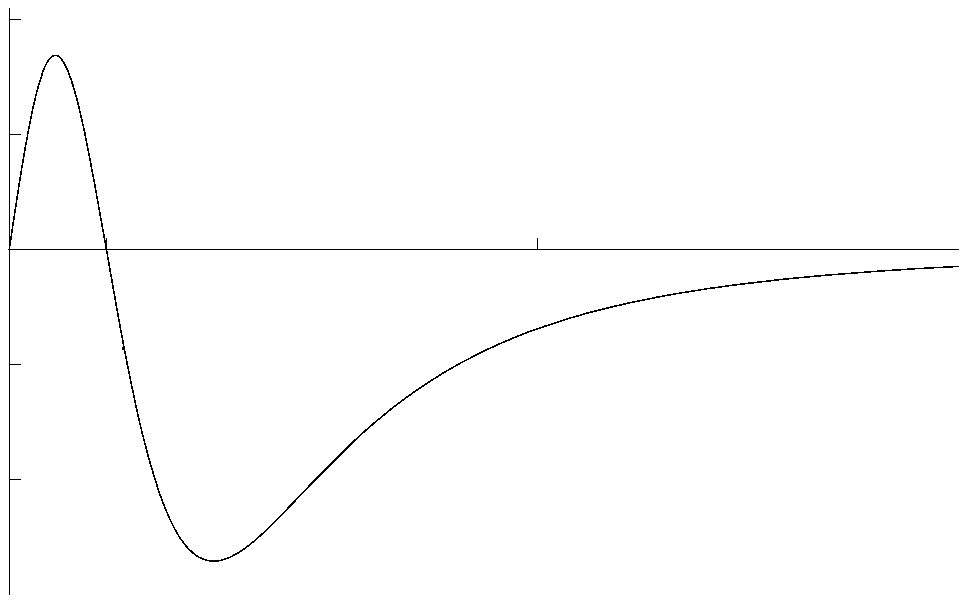}\hspace{1cm}
  \includegraphics[width=6.8cm,height=4.5cm]{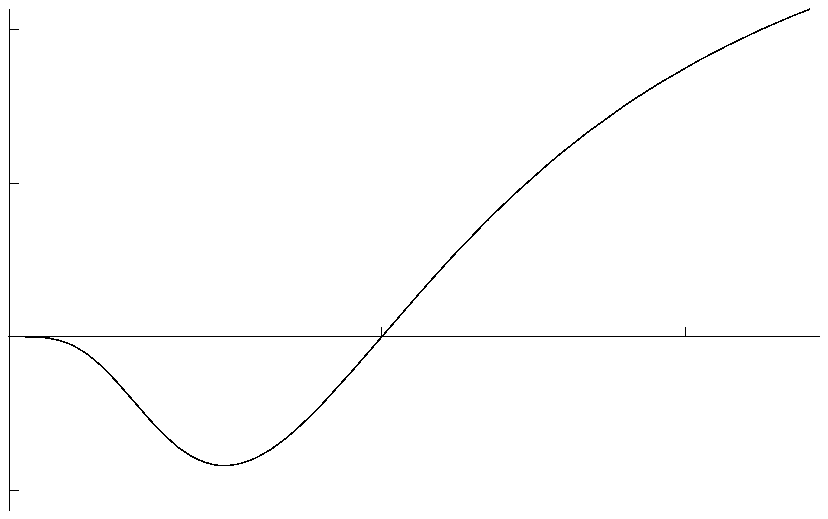}
\begin{picture}(0,0)(0,0)
\begin{tiny}
\put(-145,23){$(2\pi\sqrt{3})^{-1}$}
\put(-113,23){$0.5$}
\put(-150,43){$1$}
\put(-152,34){$0.5$}
\put(-150,25){$0$}
\put(-154,17){$-0.5$}
\put(-152,8){$-1$}
\put(-80,25){$\frac{T}{\Omega}$}
\put(-150,46){$G\, \Omega^3 C_{\Omega}$}
\put(-76,42){$0.0003$}
\put(-77,28){$0.00015$}
\put(-71,15){$0$}
\put(-79,1){$-0.00015$}
\put(-14,13){$0.5$}
\put(-39,12){$\frac{\sqrt{3}}{2\pi}$}
\put(-70,46){$G\,T^4 \epsilon_{T} $}
\put(-1,14){$\frac{T}{\Omega}$}
\end{tiny}
\end{picture}
\vspace{0.5cm}
\caption{\small On the right, the heat capacity $ C_{\Omega}$ as a function of the temperature $T$ (at a fixed value of $\Omega$) and on the left, the compressibility $\epsilon_T$ as a function of the angular velocity $\Omega$ (at a fixed value of $T$) of a five dimensional singly spinning black hole is shown. The heat capacity is positive in the region where $ (T/\Omega)\,< \left(2 \pi \sqrt{3} \right)^{-1}\simeq 0.092$ (or equivalently for $\nu > 0.6$) and the compressibility is positively defined in the region where $  (T/\Omega)>\sqrt{3}/ (2 \pi)\simeq 0.276$ (or for $\nu < 0.14$) implying the instability of the black hole in the canonical and grand canonical ensembles. The heat capacity tends to zero when approaching the singular extremal black hole solution with $T=0$.}\label{fig:BH-GranCan}
\end{figure}

A regular black hole solution with one angular momentum corresponds to
setting  $\lambda=1$ in (\ref{generalmetricunbalanced}). Thus, the Gibbs
potential is given by
\begin{eqnarray}
{\cal G}[T,\Omega]=\frac{\pi }{8 G \,\Omega^2}\left(1+\frac{4 \pi ^2 T^2}{\Omega ^2}\right)^{-1}\,,
\end{eqnarray}
and the specific heat is
\begin{eqnarray}\label{eq:heatBR}
G \Omega ^3\,C_{\Omega}=\frac{\pi ^2 x \left(1-3 x^2\right)}{2 \left(1+x^2\right)^3}\,,
\end{eqnarray}
where $x=2 \pi T /\Omega $. Thermodynamic stability, $C_\Omega>0$,
restricts $T/ \Omega <  \left(2 \pi \sqrt{3}  \right)^{-1}\simeq 0.092$
which in turns implies $\nu > 3/5 \simeq 0.6 $. Although the solution is
singular when $\nu\rightarrow 1$, in the extremal limit the heat capacity
tends to zero, $C_{\Omega}\rightarrow 0$, as shown in Fig.\ref{fig:BH-GranCan}.

The compressibility can be shown to be
\begin{eqnarray}\label{eq:compBR}
G T^4\,\epsilon_T=\frac{1-3 \bar{x}^2}{64 \pi ^3 \left(1+\bar{x}^2\right)^3}\,,
\end{eqnarray}
where $\bar{x}= \Omega/(2\pi\, T)$.
In Fig.\ref{fig:BH-GranCan} we show the compressibility as a function of the angular velocity for a fixed value of the temperature. Therefore, it is positive for $(\Omega /T )>2 \pi /\sqrt{3}\simeq 3.63$ that corresponds to a constraint on the parameters of the solution so that $\nu < 1/7 \simeq 0.14$. As for the limit of $\Omega\rightarrow 0$ corresponding to Schwarzschild black hole it is observed that the compressibility is positive.\footnote{It is well known that the heat capacity for a
Schwarzschild black hole is negative and so it heats up as it radiates (it is not thermodynamically
stable). However, since the compressibility is positive, it is stable against perturbations in
the angular velocity.}

The response functions are positive for different values of the parameters implying no region of the parameter space where both are simultaneously positive. Therefore, the black hole is thermally unstable in both, the grand-canonical and canonical ensembles.

\subsubsection{Black ring}
We consider now the dynamical equilibrium condition $\lambda=2\nu/(1+\nu^2)$ that corresponds
to a regular black ring with one angular momentum.

\begin{figure}
\centering
  \includegraphics[width=7cm,height=4.5cm]{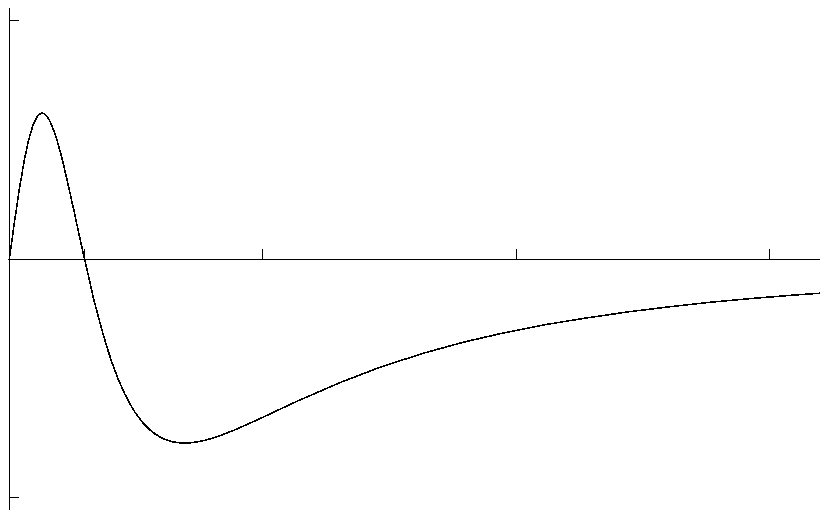}\hspace{0.5cm}
  \includegraphics[width=7cm,height=4.5cm]{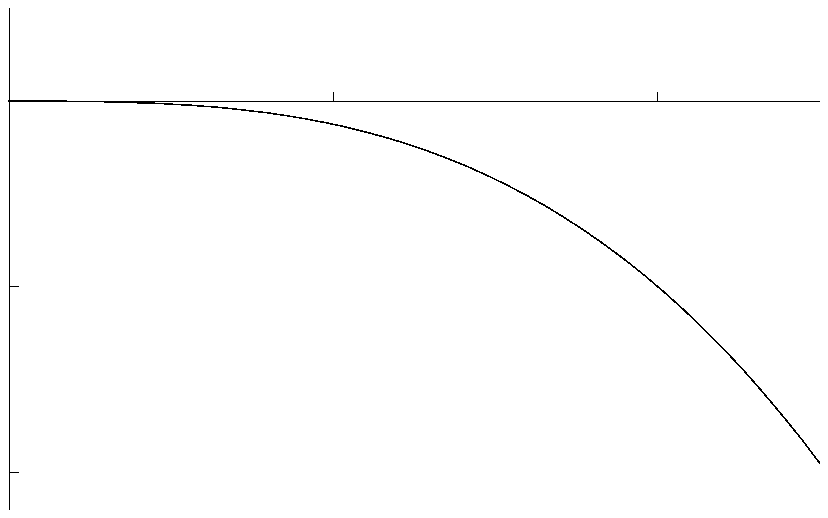}
\begin{picture}(0,0)(0,0)
\begin{tiny}
\put(-141,20){$0.074$}
\put(-126,20){$0.25$}
\put(-108,20){$0.5$}
\put(-85,20){$0.75$}
\put(-148,43){$1$}
\put(-148,21){$0$}
\put(-150,1){$-1$}
\put(-77,21){$\frac{T}{\Omega}$}
\put(-150,46){$G\, \Omega^3 C_{\Omega}$}
\put(-71,36){$0$}
\put(-74,19.5){$-1$}
\put(-74,3){$-2$}
\put(-16,33){$2$}
\put(-43,33){$1$}
\put(-74,46){$G\,T^4 \epsilon_{T} $}
\put(-1,35){$\frac{T}{\Omega}$}
\end{tiny}
\end{picture}
\vspace{0.5cm}
\caption{\small Plot of the heat capacity $ C_{\Omega}$ as a function of the temperature $T$ for a given angular velocity (left) and the compressibility $\epsilon_T$ as a function of the angular velocity $\Omega$ for a given temperature (right) for a singly spinning black ring. The black ring is unstable in the canonical and grand canonical ensembles: the compressibility is always negative and the heat capacity is only positive when $ T/\Omega\, <  \left(4 \pi \sqrt{2/ \sqrt{3}}\right)^{-1}\simeq 0.074$. In the extremal limit ($T=0$) the heat capacity of the black ring is zero.}\label{fig:BR-GranCan}
\end{figure}

The Gibbs potential can be written as a function of the temperature and the angular velocity as follows
\begin{eqnarray}\label{eq:GBR}
{\cal G}[T,\Omega]=\frac{\pi }{4 G  \, \Omega ^2}\left(1+ \sqrt{1+\frac{16 \pi ^2 T^2}{\Omega ^2}}\right)^{-1}\,.
\end{eqnarray}
A straightforward computation leads to
\begin{eqnarray}
G \Omega ^3\,C_\Omega=\frac{\pi ^2 x \left(1+\sqrt{1+x^2}-2 x^2 \sqrt{1+x^2}\right)}{\left(1+x^2\right)^{3/2} \left(1+\sqrt{1+x^2}\right)^3}\,,
\end{eqnarray}
where $x= 4 \pi T /\Omega$. It turns out that solutions with $ T/\Omega\, <  \left(4 \pi \sqrt{2/ \sqrt{3}}\right)^{-1}\simeq 0.074$ are stable
against thermal fluctuations, $C_\Omega>0$.
It is also important to note
that in the extremal limit where $T \rightarrow 0$ or $\nu \rightarrow 1$
the heat capacity goes to zero as shown in Fig. \ref{fig:BR-GranCan}. This
behavior for the heat capacity is expected and can be drawn from Nernst theorem.

Similarly the compressibility can be computed
\begin{eqnarray}
G T^4\,\epsilon_T=-\frac{2+3 \bar{x}^2}{1024 \pi ^3 \bar{x}^3 \left(1+\bar{x}^2\right)^{3/2}}\,,
\end{eqnarray}
where $\bar{x}= \Omega/ (4 \pi  T)$ and so, for any value of the parameters, it is always
negative. Therefore, the black ring is also thermodynamically unstable in the grand-canonical
and canonical ensembles.

\section{Charged black objects}
In this section we compute the stress tensor of charged $5$-dimensional black objects. In particular, we discuss the Reissner-Nordstrom black hole, a supersymmetric black ring solution, and black string solutions in Einstein-Maxwell-dilaton theories. We also obtain the equilibrium condition  for the black string solutions that are obtained as limit  of black rings in \cite{Emparan:2004wy}.

\subsection{Reissner-Nordstrom black hole in five dimensions}
As a warm-up exercise, we begin by analyzing the Reissner-Nordstrom black hole.
The static black hole solution of Einstein-Maxwell field equations has
the following line element
\begin{eqnarray}
ds^2=-V(r)\,dt^2+\frac{dr^2}{V(r)}+r^2 \,d\Omega^2_3 \,,
\end{eqnarray}
where
\begin{eqnarray}
V(r)=1-\frac{m}{r^2}+\frac{q^2}{r^4}\,.
\end{eqnarray}
The parameters $(m,q)$ are related to the mass and electric charge respectively. The coordinates range between $0\leq\theta<\pi/2$ and $0\leq\phi,\psi<\pi$.

Using the counterterm method we find the relevant stress energy component
\begin{equation}
\label{RN5D}
\tau_{tt}=\frac{1}{8 \pi G}\left(-\frac{3}{2}  \, \frac{m}{r^3}+\left(\frac{9 m^2}{8}+\frac{3 q^2}{2}\right) \,\frac{1}{r^5}+\mc{O}(1/r^7)\right)\, .
\nonumber\\
\end{equation}
As expected, the charge contribution is subleading in the $\tau_{tt}$ component of
the stress tensor.

From (\ref{charge}) we can then calculate the conserved mass associated with the closed surface $\Sigma$
\begin{eqnarray}
M \equiv \oint_{\Sigma} d^3y \sqrt{\sigma} n^i\,\tau_{ij}\,\xi_t^j=\frac{3\,m}{8\,G}\, ,
\nonumber
\end{eqnarray}
where the the normalized Killing vector associated with the mass is $\xi_t=\partial_t$,
matching the ADM computation.

\subsection{The supersymmetric black ring}
This is the solution of the bosonic sector of five-dimensional minimal supergravity
with  an action principle
\begin{eqnarray}
\label{action-sugra} I_0=\frac{1}{16\pi G}\int_M d^5x\left(\sqrt{-g}
{\left(R- F^2\right)} - \frac{8}{3\sqrt{3}}F\wedge F\wedge A\right) -\frac{1}{8\pi G}\int_{\mathcal{\partial
M}}K \sqrt{-h}\,d^{4}x ,
\end{eqnarray}
and field equations
\begin{eqnarray}
\label{eqs-sugra} R_{\mu \nu}-\frac{1}{2}Rg_{\mu \nu}=2(F_{\mu
\sigma}F^{\sigma}_{\nu}-\frac{1}{4}g_{\mu \nu}F^2),
\\
D_{\mu}F^{\mu \nu}=\frac{1}{2\sqrt{3}\sqrt{-g}} \epsilon^{\nu \mu
\sigma \lambda\tau}F_{\mu \sigma}F_{\lambda\tau},
\end{eqnarray}
where $F=dA$.

The line element of the black ring solution is given by
\cite{Elvang:2004rt} (see also \cite{Emparan:2006mm,Elvang:2004ds, Bena:2004de, Bena:2007kg})
\begin{eqnarray}
\label{metric-ss} ds^2 = -f^2(dt+\omega)^2 + f^{-1} ds^2_4 \,,
\end{eqnarray}
with  the flat space metric  written in ring coordinates
\begin{eqnarray}
ds^2_4= \frac{R^2}{(x-y)^2}  \Big[ \frac{dy^2}{y^2-1} +
(y^2-1)d\psi^2  +\frac{dx^2}{1-x^2}+(1-x^2)d\phi^2 \Big] \,.
\end{eqnarray}
where
\begin{eqnarray}
f^{-1}=1+\frac{Q-q^2}{2R^2}(x-y)-\frac{q^2}{4 R^2}(x^2-y^2),~~
\omega= \omega_{\psi}d\psi+\omega_{\phi} d\phi \, ,
\end{eqnarray}
and
\begin{eqnarray}
\omega_\phi &=& -\frac{q}{8R^2} (1-x^2) \left[3Q - q^2 ( 3+x+y)
 \right]\,, \\
\omega_\psi &=& \frac{3}{2} q(1+y)  + \frac{q}{8R^2} (1-y^2)
\left[3Q - q^2 (3 +x+y) \right]\, .
\end{eqnarray}
The gauge potential is
\begin{eqnarray}
 A=\frac{\sqrt{3}}{2} \left[ f \, (dt+\omega) -
 \frac{q}{2} ((1+x ) \, d\phi + (1+y)  \, d\psi) \right] \,.
 \label{apot}
\end{eqnarray}
The coordinates have ranges $-1\leq x\leq 1$ and $-\infty<y\leq
-1$, and $\phi,\psi$ have period $2\pi$. The black ring has an event
horizon at $y=-\infty$. $Q$ and $q$ are positive constants,
proportional to the net charge and to the dipole charge of the
ring, respectively. The electric charge relevant for
thermodynamics is ${\cal Q}=\sqrt{3}Q/2$.

The same counterterm approach can be used to compute the asymptotic
conserved charges. In this computation, it is
convenient to use the $(r,\theta)$ coordinates, defined by
(\ref{transf}) with $\alpha=1$. The relevant components of the
boundary stress tensor are
\begin{eqnarray}
\label{susyBRtensor}
\tau_{tt}&=&\frac{1}{8 \pi G}\left(-\frac{3 Q}{r^3}-\frac{5}{3} R^2 \frac{\cos 2 \theta }{r^3}+\mc{O}(1/r^5)\right)\,,\nonumber\\
\tau_{t\phi}&=&\frac{1}{8 \pi G}\left(-q (3Q-q^2) \frac{\cos^2 \theta }{r^3}+\mc{O}(1/r^5)\right)\,,\nonumber\\
\tau_{t\psi}&=&\frac{1}{8 \pi G}\left(-q (6\,R^2+3Q-q^2) \frac{\sin^2 \theta}{r^3}+\mc{O}(1/r^5)\right)\,,\nonumber\\
\tau_{\theta\theta}&=&\frac{1}{8 \pi G}\left(\frac{2}{3} R^2 \frac{\cos 2 \theta }{r}+\mc{O}(1/r^3)\right)\,,\\
\tau_{\phi\phi}&=&\frac{1}{8\pi G}\left(\frac{2}{3}R^2(1+2\cos2\theta)\frac{\cos^2\theta}{r}+\mc{O}(1/r^3)\right)\,,\nonumber\\
\tau_{\psi\psi}&=&\frac{1}{8 \pi G}\left(\frac{2}{3}R^2(-1+2\cos2\theta)\frac{\sin^2\theta}{r}+\mc{O}(1/r^3)\right)\,.\nonumber
\end{eqnarray}
Therefore, the mass and angular momentum as computed from the
counterterms are the same as the ADM values
\begin{eqnarray}
\nonumber
M=\frac{3\pi Q}{4G},~~J_\varphi=\frac{\pi}{8G}q(3Q-q^2),~~
J_\psi=\frac{\pi}{8G}q(6\,R^2+3Q-q^2)\, .
\end{eqnarray}

For this supersymmetric solution, the surface gravity and the
angular velocities of the event horizon vanish. Despite this,
the horizon area is finite and depends on both the global and
dipole charges. We present more details on the role of the charges and
the thermodynamics of the supersymmetric black ring in the Discussion
section.

\subsection{Black string and balance condition}
In this section we discuss the 5-dimensional charged boosted black string
solutions in Einstein-Maxwell-dilaton theory \cite{Emparan:2004wy} by using
the counterterm method. The action is
\begin{equation}\label{emdaction}
I=\frac{1}{16\pi G}\int d^5x\sqrt{-g}\left(
R-\frac{1}{2}(\partial\phi)^2-\frac{1}{4}e^{-\alpha\phi} F^2\right)\,.
\end{equation}

It is convenient to express the dilaton coupling as in
\cite{Emparan:2004wy}
\begin{equation}
\label{alphan}
\alpha^2=\frac{4}{N}-\frac{4}{3}\,,\qquad 0< N\leq 3\,.
\end{equation}

We will obtain
the relevant components of the stress tensor and discuss the balance
condition.\footnote{A more detailed discussion of the equilibrium
condition for thin neutral black rings within the quasilocal formalism
and also the generalization to `fat' black ring solutions was presented
in \cite{Astefanesei:2009mc}.}

The solution, with the boost parameter $\sigma$ and the event horizon $r=r_0$, is
\begin{eqnarray}
ds^2=-\frac{\hat{f}}{h^{N/3}}\left(dt-\frac{r_0\sinh \sigma \cosh \sigma}{r\hat{f}}\,dz \right)^2
+\frac{f}{h^{N/3}\hat{f}}\,dz^2
+h^{2N/3}\left(\frac{dr^2}{f}+r^2d\Omega_2^2\right)\,,
\end{eqnarray}
where the (magnetic) charge is parametrized by $\gamma$.

The gauge potential and the dilaton for the
magnetic\footnote{The expressions for the two-form
potential and dilaton of the dual electric solutions
are given in \cite{Emparan:2004wy}.}
solution are, respectively,
\begin{eqnarray}
A_{\phi}=\sqrt{N}\,r_0\sinh\gamma \cosh\gamma \,(1+\cos\theta)\,,\qquad e^{\phi}=h^{N\alpha/2}\, ,
\end{eqnarray}
and
\begin{eqnarray}
f=1-\frac{r_0}{r}\,,\qquad\hat{f}=1-\frac{r_0\cosh^2\sigma}{r}\,,\qquad
h=1+\frac{r_0\sinh^2\gamma}{r}\,.
\end{eqnarray}

The black string solutions that are obtained as limit of black rings should also satisfy
the equilibrium condition. The equilibrium condition is a constraint on the
parameters of the unbalanced ring solution (\ref{generalmetricunbalanced}) that
is equivalent with the removing of all conical singularities in the metric.

A nice physical interpretation was given in \cite{Emparan:2004wy}: the absence
of conical singularities is equivalent with the equilibrium of the forces acting
on the ring. A black ring can be obtained by bending a {\it boosted} black string.
Thus, the linear velocity along the string becomes the angular velocity of the
black ring. The equilibrium of centrifugal and gravitational forces imposes
a constraint on the radius of the ring $R$, the mass, and the angular momentum. In
this way one can see that, indeed, just two parameters are independent in the
solution of the neutral black ring.

Applying the same procedure as before we find that
the relevant component of the boundary stress tensor is
\begin{equation}
\tau_{zz}=\frac{1}{8 \pi G}\left(\frac{r_0^2}{2} \left(1-\sinh^2{\sigma}+N
\sinh^2\gamma\right)\frac{1}{r^2}+\mc{O}(1/r^3)\right)\,.
\end{equation}

From a more general definition of Carter \cite{Carter:2000wv}, in the absence
of external forces the equations of motion of brane-like objects obey $K^{\rho}_{\mu\nu}T^{\mu\nu}=0$, which implies the component of the stress
tensor in the z-direction (the pressure) vanishes $T_{zz}=0$.

Thus, asymptotically, this equality (at first order) constrains the values of the boost parameter with the charge (parameterized by $\gamma$) in the following manner
\begin{eqnarray}
\sinh^2\sigma=1+N\sinh^2 \gamma\,,
\end{eqnarray}
that is in agreement with the regularity constrains
in \cite{Emparan:2004wy} and the equilibrium condition
found in \cite{Emparan:2007wm} for thin black rings.

By direct integration of the stress energy components the conserved
charges, mass and angular momentum, match exactly those from the ADM
definition, namely
\begin{eqnarray}
M=\frac{\pi}{8G} \,R\,r_0 \cosh 2 \sigma ,\qquad J=-\frac{\pi }{8 G}\,R^2 r_0 \cosh \sigma  \sinh \sigma\, .
\end{eqnarray}

\section{Discussion}
\label{discuss}
In this paper we systematically have applied the counterterm method
for asymptotically flat spacetimes to 5-dimensional black objects. In
this way, we have derived various thermodynamic relations for
several stationary black objects.

We hope that our unified treatment of black holes, black rings, and
black strings with an emphasis on the role of the quasi-Euclidean
section for thermodynamics is useful to the reader.

\subsubsection*{On the Quasi-Euclidean method}

The notion of asymptotic flatness of isolated systems is intimately
related to the possibility of defining the total stress-energy tensor
that characterizes the gravity system \cite{Arnowitt:1962hi}. It is
well known that a spacetime is asymptotically flat if it is
possible to attach to its corresponding manifold a boundary in
{\it null} directions ($\cal{I}$) --- since null rays reach ($\cal{I}$) for an
infinite value of their affine parameter, this is called
{\it null infinity}. Spatial infinity, $\iota^0$ (the part of infinity
that is reached along spacelike geodesics), is represented by one
point in the Penrose diagram of conformal compactification for Minkowski space.

It is important to emphasize that it is also possible to foliate the
spacetime with {\it spacelike} foliations:  spacelike surfaces
can be constructed that extend through null infinity. Such surfaces
are called {\it hyperboidal} as their asymptotic behaviour is similar to
standard hyperboloids in Minkowski spacetime. In this context, it is
better to visualize $\iota^0$ as the hyperboloid of spacelike
directions (it is isometric to the unit $4$-dimensional de Sitter space).

{}From a physical point of view, $\iota^0$ can be interpreted as the
place where an observer ends up when shifted to larger and larger distances.
However, for studying holography in {\it stationary} flat spacetimes,
it seems more natural to impose boundary conditions at $\iota^0$ rather
than $\cal{I}$.\footnote{A nice discussion on the role of conformal
boundary and boundary conditions for holography can be found
in \cite{Marolf:2006bk}.} Therefore, the renormalized `boundary'
stress tensor we used in this paper is assigned to
spatial infinity $\iota^0$ of asymptotically flat spacetimes.

A key point is that one's intuition about Euclidean
sections does not apply to black rings --- there is no real {\it
non-singular} Euclidean section in this case. Therefore, a
black ring should be described by a {\it complex} Euclidean
geometry \cite {Brown:1990fk} and its associated {\it real} action
(`thermodynamic action').\footnote{A real Euclidean metric
associated with the vacuum Kerr black hole was obtained by
supplementing the analytic continuation $t\rightarrow -i\tau$ by a
further transformation in the moduli space of the parameter space,
$J\rightarrow iJ$. However, as argued by some authors
\cite{Brown:1990fk}, the resulting metric has little to do with the
physical (Lorentzian) Kerr black hole.}

Consequently we employed the quasi-Euclidean method \cite{BoothMann,Booth:1998pb}, which was applied to black rings for the first
time in \cite{Astefanesei:2005ad}. The properties of the black ring
interior become encoded in a set of conditions at the `bolt' of the
complex geometry. Thus, the partition function
computed as a functional integral is extremized by a certain
stationary complex black ring metric.

A natural question that arises is if the physical system described
by this partition function is a real Lorentzian black ring. The answer
is yes, precisely because the stress tensors for the complex black
ring and for the related Lorentzian black ring coincide. For example,
in the zero loop approximation, the expectation value of energy
from the partition function will coincide with the energy of the
complex black ring calculated from the boundary stress-tensor; in
turn, the latter characterizes the energy of the real Lorentzian
black ring.

Further support for using  quasi-Euclidean instantons to construct
gravitational partition functions was given in \cite{Monteiro:2009tc}. In
this work, the authors discuss the thermodynamic instabilities of several
spinning black objects in the grand-canonical ensemble. They found that
the partition functions of neutral spinning black holes and black rings in
flat spacetime possess negative modes at the perturbative level.

A central result in our work is the computation of the {\it renormalized}
action (\ref{quasi-Euc-action}) on the quasi-Euclidean section. The black ring solutions
have been shown to satisfy the first law of
black hole mechanics, thus suggesting that their entropy is one
quarter of the event horizon area. We have made this more
precise by computing the gravitational action to check the quantum
statistical relation as well as the first law of thermodynamics. Our
computation can be considered as an independent check that the
entropy/area relation applies also for the black rings.

\subsubsection*{On the thermodynamics}

The thermodynamics of black rings in different ensembles has been previously
presented in the literature \cite{Astefanesei:2005ad, Elvang:2006dd,
Monteiro:2009tc}. For completeness, we also present a discussion
of thermodynamic stability within quasilocal formalism.

Four-dimensional black holes are highly constrained objects. That
is, an {\it isolated} electrovac black hole can be characterized,
{\it uniquely} and {\it completely}, by just three macroscopic
parameters: its mass, angular momentum, and charge. Thus, all
multipole moments of the gravitational field are radiated away
in the collapse to a black hole, except the monopole and dipole
moments \cite{Townsend:1997ku} --- they cannot be radiated away
because the graviton has spin $2$.

There are no black objects with an electric dipole in four dimensions.
The black holes have `smooth' horizons (there are no ripples or higher
multipoles) and are classically stable. Moreover, for asymptotically
flat solutions, the event horizons of non-spherical topology are forbidden.

For vacuum Einstein gravity in more than four dimensions there is no
uniqueness since a richer range of regular \textit{black objects}
inhabit the space. These include not just black holes \cite{Myers:1986un}
with spherical $S^{D-2}$ horizon geometry, but also black rings
\cite{Emparan:2007wm} with $S^1\times S^{D-3}$ and blackfolds \cite{Emparan:2009cs}
with topological, i.e. $S^{p+2}\times S^{D-p-4}$ when $p \leq D-7 $, horizon geometries.

Many studies on the thermodynamics (some reviews \cite{Emparan:2006mm,Emparan:2008eg,Obers:2008pj,Kleihaus:2007kc}),
ergoregions \cite{Elvang:2008qi}, and combinations of black objects
leading to more sophisticated solutions
\cite{Iguchi:2007is} uphold
the exciting richness of  black objects. And moreover, as we present
here, new insights into the thermodynamics remain to be unveiled.
Perhaps because of the tight contact with string theory, the most widely
employed scheme to explore the thermodynamics of black holes in five
\cite{Emparan:2001wn,Elvang:2007hg} and higher space-time dimensions
\cite{Emparan:2007wm} seems to be the microcanonical ensemble.

In this paper, to study the stability of five dimensional black holes,
we have carried out the thermodynamic analysis mainly in the canonical and
grand canonical ensembles. Not only the thermodynamic stability but
also the phase structure depends on the chosen ensemble.

\begin{figure}[t!]

 \centering
\hspace{-0.5cm}  \includegraphics[height=5.5cm,width=6.5cm]{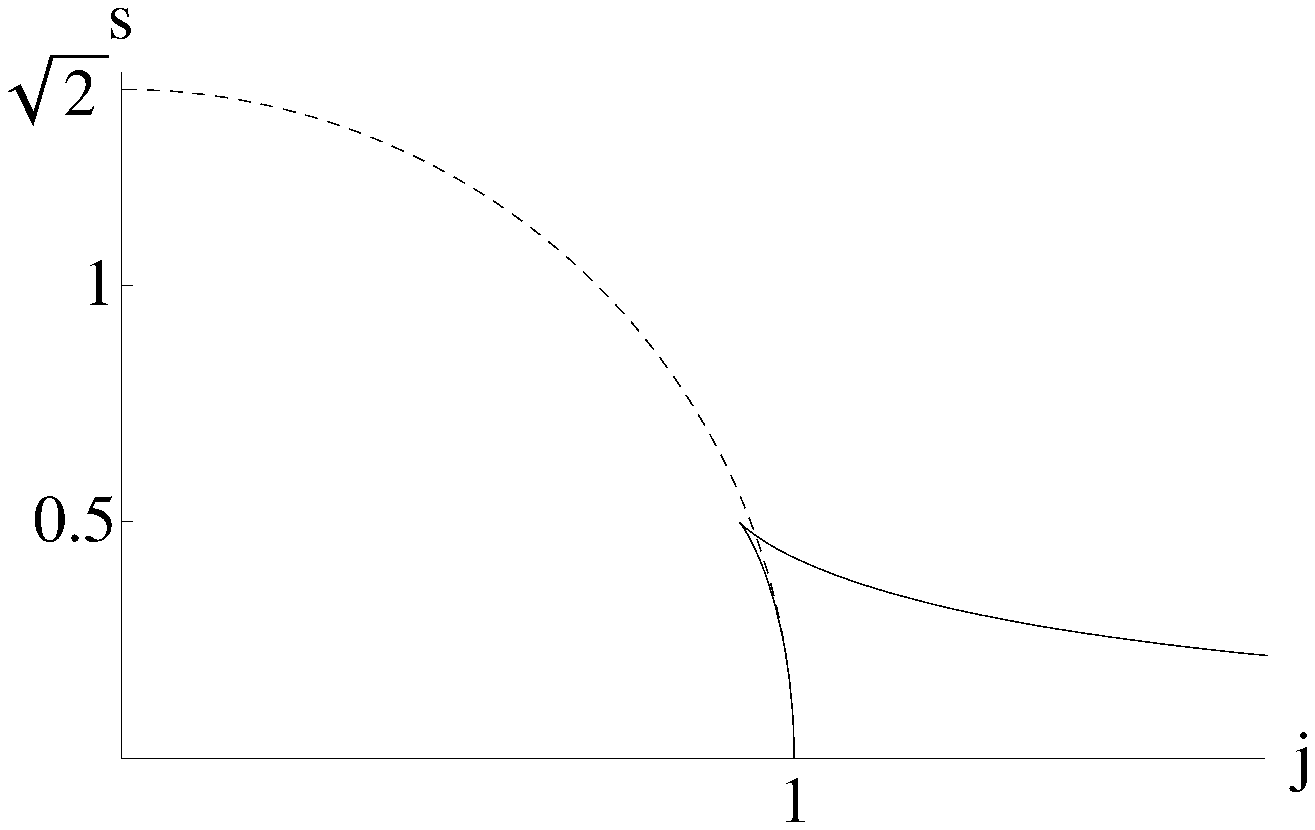}\hspace{0.4cm}
           \raisebox{0.1cm}{\includegraphics[height=5.5cm,width=6.2cm]{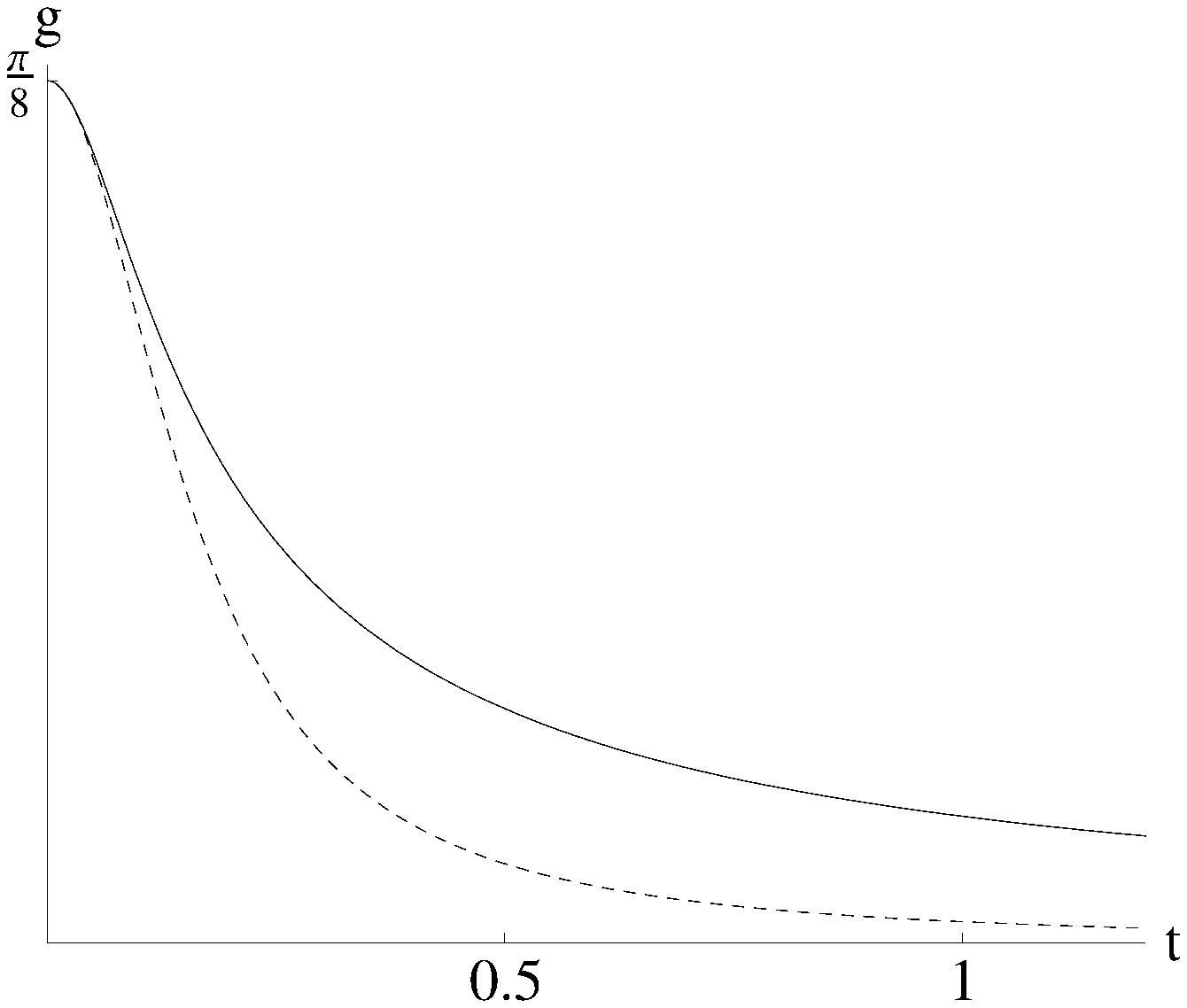}}\hspace{2.5cm}
    \includegraphics[height=5.5cm,width=6cm]{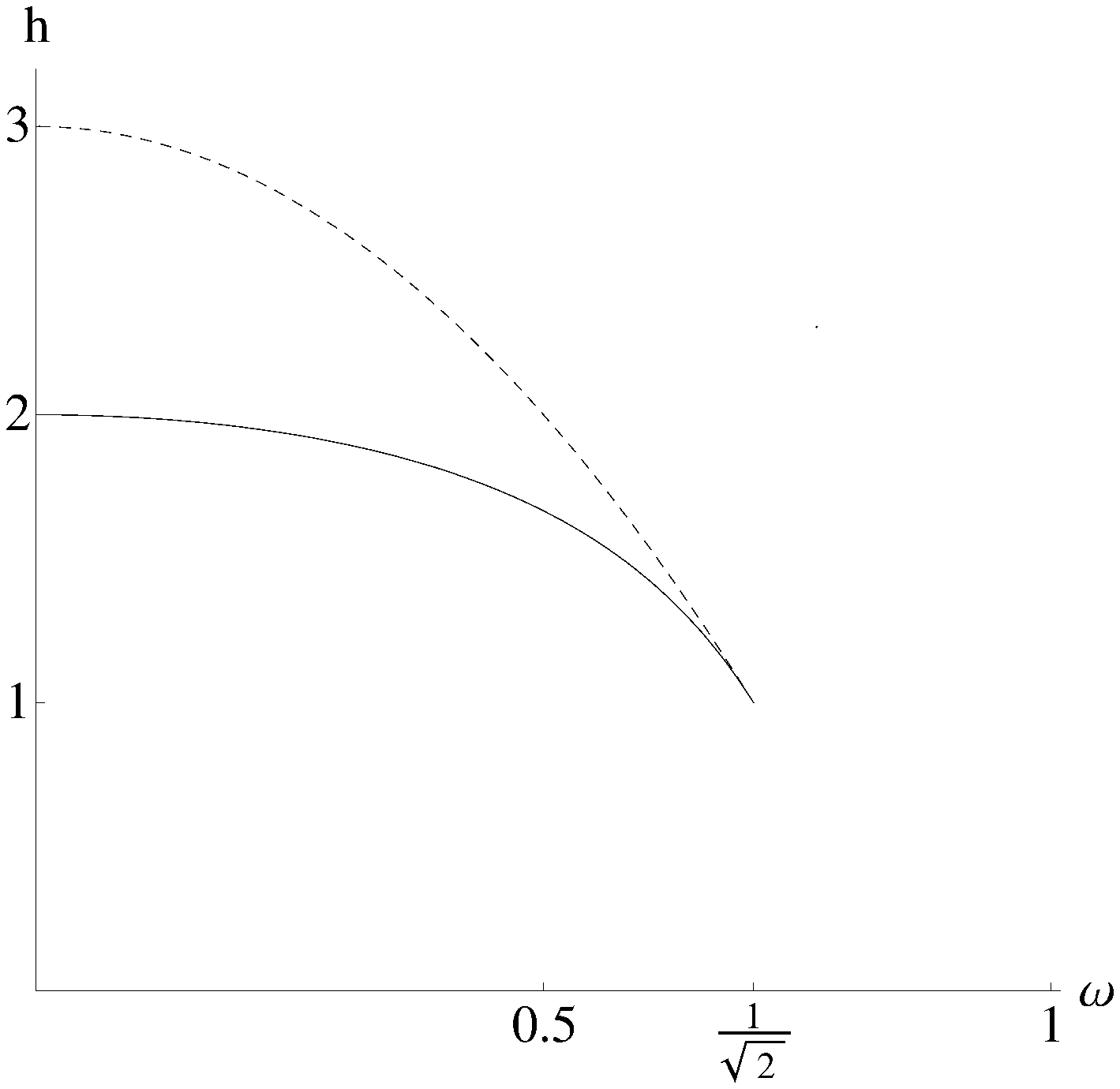}\hspace{0.6cm}
  \includegraphics[height=5.5cm,width=6cm]{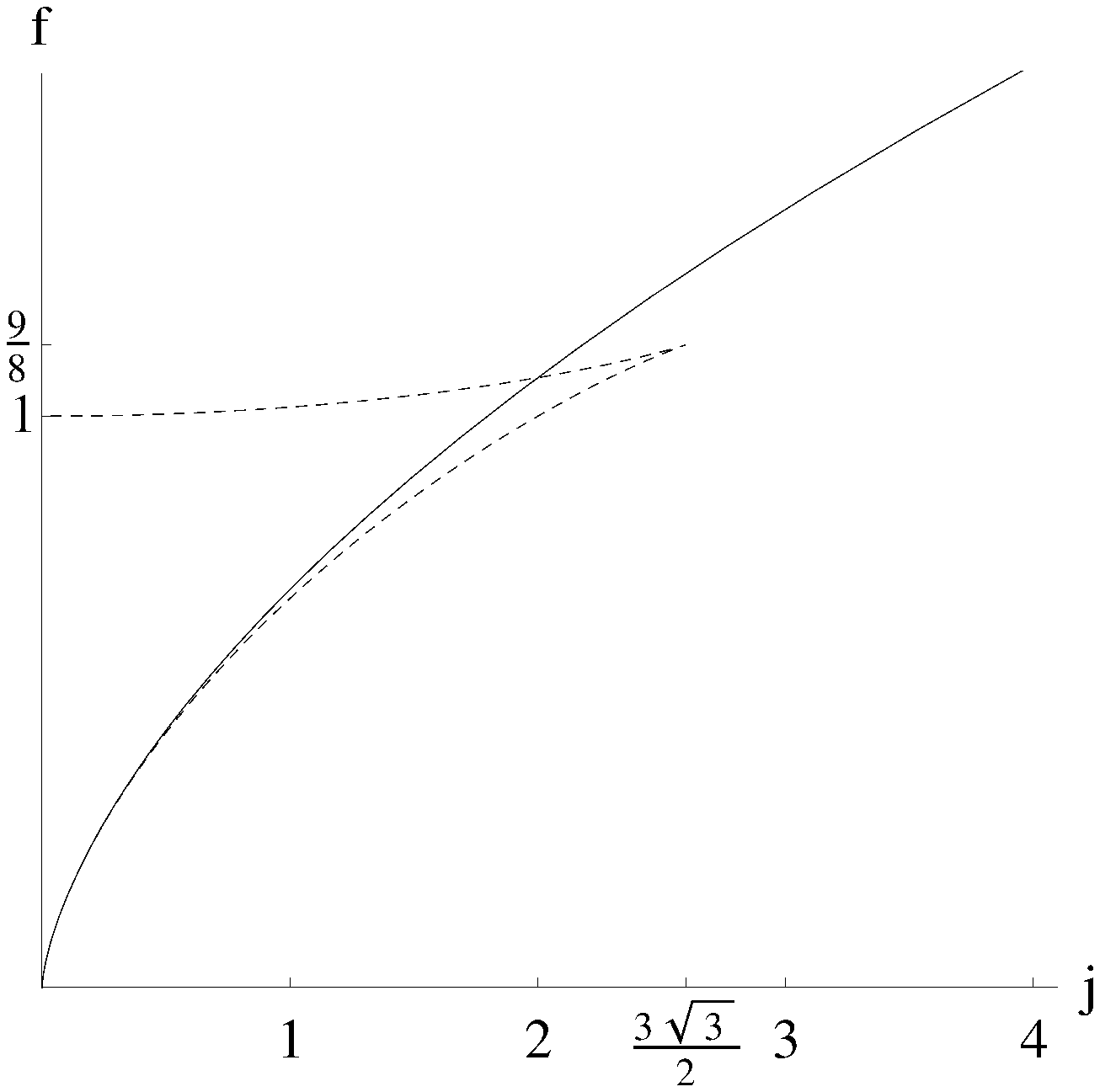}

\begin{picture}(0,0)(0,0)
\put(-105,160){(a)}
\put(-105,3){(b)}
\put(82,160){(c)}
\put(82,3){(d)}
\end{picture}

\bigskip
\caption{\small The phase diagrams of a singly spinning black hole (dashed) and a black ring (solid) in different thermodynamic ensembles: (a) microcanonical, (b) enthalpy, (c) grand canonical and (d) canonical.}\label{fig:ensembles}
\end{figure}

Already in our case,
of the canonical $f$ vs $t$ and grand canonical $g$ vs $t$ ensembles,
some significant differences can directly be noticed from the structure
of these phase diagrams\footnote{We use (\ref{eq:mandj}), (\ref{angularvel}) and (\ref{areatemp}) for the black hole/ring to define dimensionless reduced quantities for the plots. In the microcanonical ensemble, for a fixed value of the mass, the entropy is defined as $s=\frac{3\sqrt{3}}{8\,\pi}\frac{S}{\sqrt{G_5\,M^{3}}}$ and the angular momenta $j=\frac{27 \pi }{32\,G_5}\frac{J^2}{M^3}$. In the grand canonical ensemble, for a fixed value of the angular velocity, the Gibbs potential is defined as $g=G_5 \Omega^2 {\cal G}$ and the temperature as $t=T/\Omega$. In the canonical ensemble, for a fixed value of the temperature, the  free energy is defined as $f=32 \pi G_5\,T^2\,F$ and the angular momenta as $j=256 \pi ^2 G_5 J \,T^3$. And in the diagram for the enthalpy, for a fixed value of the mass, $h=H M$ and the angular velocity as $\omega=\frac{2}{\sqrt{3 \pi }}\sqrt{G_5 M}\,\Omega$. } for the black hole (dashed line) and the black ring (solid line) in Fig. \ref{fig:ensembles}.

The single phases, one for each, of the black hole and ring in the
grand-canonical ensemble contrast the three phases, two for the black hole and a single one for the black ring, of the canonical ensemble. The entropy $S[M,\Omega]=A/(4\,G)$ and the enthalpy, $H[M,\Omega]=M-\Omega\,J$ are also shown for comparison. For a fixed mass, in the microcanonical ensemble a swallowtail structure is found for the two black ring phases with a single phase for the black hole. Leaving the mass fixed, yet a different structure is found: for the $h$ vs $\omega$ each of the single black hole and black ring phases join at a maximum value of the angular velocity. The plots we present here are new although the discussions comparing the different structures of the phase diagrams can also be found in \cite{Elvang:2006dd}.

\subsubsection*{On the nature of charges}

Another case where the classical uniqueness results do not apply is
for gravity theories with scalar fields non-minimally coupled to gauge fields.
Due to the non-minimal coupling, a black hole solution in
Einstein-Maxwell-dilaton gravity can carry also a scalar charge and
the first law gets modified \cite{Gibbons:1996af}.\footnote{However,
this result should be taken with caution: in string theory
the asymptotic values of the moduli `label' different ground states
(vacua) of the theory and so it is necessary an infinite amount of
energy to change the state of the system in this way --- see \cite{Astefanesei:2006sy} for a more detailed discussion.}

The scalar charge is not protected by
a gauge symmetry and so it is not conserved. Since it depends on the
other conserved charges it does not represent, though, a new quantum
number associated with the black hole --- this kind of charge is called
{\it secondary hair}. Furthermore, this charge is not {\it localized} and exists
outside the horizon.

In Section $4$ we  studied a supersymmetric ring solution in minimal supergravity. Since it is an extremal black object a computation of the
action is not possible (the periodicity of
the `Euclidean time' cannot be fixed). However, one can compute the entropy by using the
entropy function formalism for spinning extremal black holes \cite{Astefanesei:2006dd}.

Due to the attractor mechanism the entropy does not depend on scalar
charges, but it depends on both, the global charge and the {\it dipole}.
Since the dipole is a non-conserved charge, one may be tempted to make
an analogy with the scalar charge. However, there is an important
difference: unlike the scalar charge, the dipole charge is a {\it localized} charge.
Thus, it can be measured by flux integrals on surfaces linked
to the black ring's horizon \cite{Emparan:2004wy} and has a microscopic
interpretation (brane wrapping contractible cycles of CY).

One important question emerges: what is the interpretation of the
dipole within the quasilocal formalism? In other words, can an asymptotic
observer distinguish between a black hole or a black ring with the same
conserved charges? The answer is obviously yes: analogous to an electric
dipole whose moments can be read off from a multipole expansion at infinity,
the subleading terms in the boundary stress tensor should encode the
information necessarily to distinguish between black objects with
different horizon topologies in the bulk.\footnote{In the supersymmetric
case, unlike the black ring, the black hole should have both angular
momenta equal. Therefore, an asymptotic observer has to compare just the angular
momenta to find out what is in the bulk.}

There is, though, another subtlety we would like to discuss in detail
now. Due to the existence of the Chern-Simons interaction in the Lagrangian,
the equation of motion for the gauge field is modified. Therefore, the
topological density of gauge field itself becomes the source of electric
charge. Consequently, even if it is conserved, the usual Maxwell charge
of a black ring seems to be diffusely distributed throughout the
spacetime. Thus, the `Maxwell charge' in this case is gauge invariant and
conserved but not localized.

The resolution of this problem was provided in \cite{Arsiwalla:2008gc,
Hanaki:2007mb}: the correct charges
that appear in the entropy are the `Page charges'. These charges are conserved and
localized but not gauge invariant (see \cite{Marolf:2000cb,Compere:2009iy} for a discussion of different
kinds of charges). It was shown in \cite{Arsiwalla:2008gc} that once the entropy function is
expressed in terms of these physical 5-dimensional charges it becomes manifestly
gauge invariant (due to a spectral flow symmetry of the theory).

The near horizon geometry of a black hole should capture the complete information
about its {\it microscopic degeneracy}.
However, if a black hole does have `hair' (degrees of freedom living outside
the horizon), there are subtle distinctions between the asymptotic charges
and the charges entering in the CFT \cite{Banerjee:2009uk}.

In the case of the susy black ring, one angular momentum is generated by degrees
of freedom living outside the horizon in the form of crossed electric and magnetic
supergravity fields \cite{Elvang:2004ds, Bena:2005zy}.

However, for the supersymmetric black ring the microscopic angular momentum
density is not equal to the angular momentum at infinity.
Interestingly enough, the `Page' angular momentum \cite{Hanaki:2007mb}
is in fact the {\it intrinsic} angular momentum of the susy black ring
and our arguments support the point of view in \cite{Bena:2004tk}.\\

\noindent While our discussion has focussed on   stationary solutions, it will
be interesting to investigate whether similar methods can be useful as well
for studying the time-dependent backgrounds given in \cite{Tasinato:2004dy}.
These solutions are obtained by a simple analytic continuation of a black
hole geometry. At a first look, it seems that the stress tensor of
these time dependent solutions should be somehow related to the
stress tensor of the `seed' black hole solution. However,
this case is more subtle since the energy-momentum carried away by
gravitational radiation is associated to null infinity.

Finally, we want to mention that the counterterm method can be also useful in
investigating the thermodynamics of the black rings obtained in
\cite{Yazadjiev:2005aw} (e.g., \cite{art}).

\section*{Acknowledgements}
We would like to thank Eugen Radu and Stefan Theisen for collaboration on
related projects and valuable discussions. DA would also like to thank
Xerxes Arsiwalla, Jan de Boer, Kostas Skenderis, and Marika Taylor for
interesting conversations and the University of Amsterdam for hospitality.
This work was supported in part by the Natural Sciences and Engineering
Research Council of Canada.

\end{document}